# Remote Oscillatory responses to a solar flare

A. Andić [1], R.T.J. McAteer[1]
Astronomy Department, MSC 4500, NMSU, P.O. Box 30001, Las Cruces, NM 88003, USA

**ABSTRACT**

The processes governing energy storage and release in the Sun are both related to the solar magnetic field. We demonstrate the existence of a magnetic connection between energy released caused by a flare and increased oscillatory power in the lower solar atmosphere. The oscillatory power in active regions tends to increase in response to explosive events at a different location, but not in the region itself. We carry out timing studies and show that this is probably caused by a large scale magnetic connection between the regions, and not a globally propagating wave. We show that oscillations tend to exist in longer lived wave trains at short periods (P< 200s) at the time of a flare. This may be a mechanism by which flare energy can be redistributed throughout the solar atmosphere.

Subject headings: Sun: photosphere, Sun:oscillations, Sun:magnetism

## 1. Introduction

Solar eruptive events (in the form of flares and CMEs) require about $10^{32}$ ergs of energy to be built up over the order of hours, and released over the order of minutes. The mechanism by which this energy is created, supplied to the solar atmosphere, and stored remains a topic of much research (e.g. Wedemeyer-Böhm et al. 2012, Kilmchuk 2012, Huang et al. 2012, Song &Vasyliunas 2011, DePontieu et al. 2009). Wedemeyer-Böhm et al. (2009) show that one of the issues with this research derives from the interconnection of photospheric oscillations and the magnetic field in the photosphere, chromosphere and corona. This interconnection may bring new insight to the issue of energy supply and storage for the solar atmosphere.

A strong connection exist between the photosphere and upper atmospheric layers. Rosenthal et al. (2002) modeled oscillations in the magnetic solar atmosphere and found that an oscillatory mode conversion occurred in the presence of magnetic field, hinting that the relative alignment between the magnetic field **B**, the wave-vector **k**, and change in plasma beta $\nabla\beta$ (ratio between magnetic and plasma pressure) might be important control parameters. Bogdan et al. (2003) found that oscillations have a complex behavior that depends on several factors: the location of the wave source; the observation line of sight that influences how we see the character of the oscillations; the position of the thin layer where sound speed and the Alfvén velocity have comparable magnitude; the number of the observed wave trains; the propagation direction of the individual wave trains; the type of plasma in which oscillations are observed. Aspects of this theoretical work were confirmed by Bloomfield et al. (2006), who observed variable wave-speeds as a wave moved from the quiet Sun network into plage and umbrae, indicative of a transition from the dominant fast-magnetoacoustic wave to the slow mode. McAteer et al. (2005) detected oscillations of period 40-80 s along a flare ribbon during the impulsive phase of a flare.

They concluded that measured properties of those oscillations were consistent with the existence of flare-induced acoustic waves.

The unveiling of this complex relationship between oscillations and the magnetic field continued with the work of Shelyag et al. (2009), who showed that the presence of a strong magnetic field perturbs and scatters acoustic waves and absorbs the acoustic power of the wave packet. Khomenko et al. (2008a) found that oscillatory events can be generated by horizontal motions of flux tubes in the photosphere and this was subsequently confirmed through the observational detection of a significant oscillatory signal that was co-spatial and co-temporal with bright points (Andic et al. 2010). These authors noted an increase of oscillatory power when bright points, (i.e. foot-points of flux tubes) are resolved and followed across the field of view (FOV). De Pontieu et al. (2004) demonstrated that propagating oscillations are channeled upward via inclined small-scale magnetic flux tubes. Khomenko et al. (2008b) broadened this research demonstrating that radiative losses might allow propagation through non-inclined small-scale magnetic flux tubes as well. Vigeesh et al. (2009) modeled upward propagation of the oscillations in flux tubes with similar behavior as in previous studies, and subsequently determined that granulation buffeting tends to produce the stronger oscillations than vortex-like motion (Vigeesh et al. 2012). de Wijn and McIntosh (2010) provided an observational work where they observed a preference in wave leakage at the edge of a plage region.

Analysis of data from the Solar Dynamics Observatory (SDO; Pesnell et al. 2012) provided a new insight into this complex interconnection between solar atmospheric layers. Schrijver & Title (2011) used SDO data to establish that the magnetic field is globally interconnected by analyzing an example where one flare triggered multiple others. Mikic et al. (2011) and Titov et al.

(2011) presented a theoretical explanation of these events. Their study showed that flares do not just change the local magnetic field, but affect the global magnetic field. These authors speculate that it is possible that a sympathetic trigger is carried by the magnetic field from one region to another.

Oscillations in solar prominences and filaments can be excited by distant flares (Ramsey & Smith 1966). Triggers of this observed filament activation are thought to be Moreton or EUV waves produced from remote flares (Balasubramaniam et al, 2007, 2010, Asai et al. 2012, Li & Zhang 2012). Jackiewicz & Balasubramamniam (2013) found recently that such a reaction can be observed even when there is no detectable Moreton wave.

Our study focuses on flares and the disturbances they cause, in order to quantify whether changes in the magnetic field can cause a visible increase in the oscillatory flux. Section 2 presents the methods used in this study, Section 3 presents an overview of the results from several different regions. Section 4 contains a detailed discussion of the results, and then implications for the current state of the knowledge in the area, with recommendations for future research directions.

## 2. Methods

The data used in this study were obtained with the Atmospheric Imaging Assembly (AIA, Lemen et al. 2012) onboard SDO on December 25th, 2011 in the 1700 ˚A passband. The dataset covers a period of 20 hours, during which 7 explosive events occurred, 6 C class flares and one M flare (Table 1). Two of these were located in region AR 11386, while the remaining five were located in emerging region AR 11387.

We analyzed five different regions on the Sun, each encompassing an area of 480 by 480 arcsec (Fig. 1). Two of areas were located in quiet Sun regions, while three encompassed active regions. Regions AR 11383 and AR 11384 were located in a single area. The area that encompassed AR 11385 did not include the location of newly emerging AR 11387.

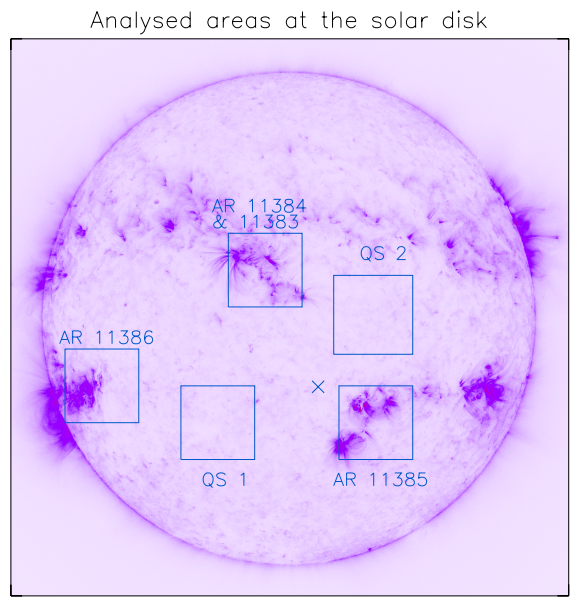

Fig. 1.— Areas used in this study. Squares represent five different areas of the solar disk we choose to analyze. Blue x marks the location of the emerging AR 11387.

The data were formed into a series of cubes, each of which spanned 40 minutes. The frames in each individual cube were corrected for solar rotation and afterwards co-aligned by a Fourier routine. This routine uses cross-correlation techniques and squared mean absolute deviations. The procedure was iterated four times to achieve sub-pixel co-alignment accuracy. This procedure can only be carried out accurately for images where clear structures are visible, a condition which is fulfilled by this dataset (Andic 2007, Jess et al. 2007).

Each light curve within each cube was searched for oscillations via a wavelet analysis (Torrence & Compo,1998), using the automatic method described in detail in Andic et al. (2010). A randomization test (Bloomfield et al., 2006) was used during the time when no explosive events were present to eliminate all power detections that were weaker than 10% of the maximum signal. We limited oscillations to those with periods greater than 30 seconds and shorter than 16.67 minutes. We also limited detections to those oscillations with at least four full cycles, and that sustained power above 20% of the maximum signal for at least 1.5 cycles.

## 3. Results

Andic et al (2010) showed that when the source of an oscillation is not resolved in the data, the oscillatory signal may be lost due to the shift of the source itself across the FOV.

as the signal stopped in one pixel and started in another. One way to eliminate the influence of the movement of the source is an integration of the oscillatory power over the whole FOV. Of course, this procedure will only show a result when a significant percentage of the oscillatory power is caused by a movement of flux tubes in the FOV, as predicted in Khomenko et al (2008b). Oscillations generated in such a manner would reach the upper layers of the photosphere and the chromosphere (De Pontieu et al. 2004; de Wijn and McIntosh, 2010). Our study also demonstrates that a large flare can result in a significant increase

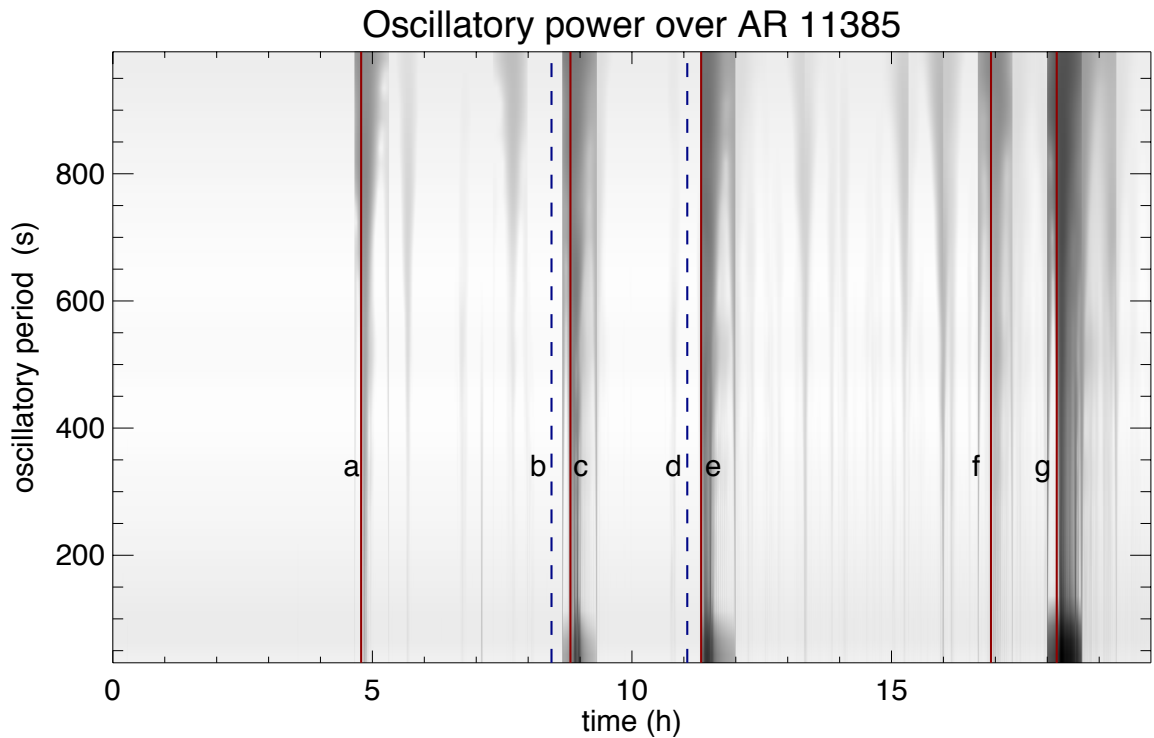

Fig. 2.— Result of the wavelet analysis of first 20 hours of the SDO dataset in 1700 centered on AR 11385 obtained on December 25, 2011. The five red solid vertical lines mark times of flares associated with emerging active region AR 11387, while the two blue dashed vertical lines mark times of the flares from AR 11386. Each flare is marked with a letter corresponding the same letter in Table 1. The increase of 90% of detected oscillatory power corresponds well with the time of the flares connected with AR 11387. No increase of the oscillatory power was registered for the flares from AR 11386. Power in this image is represented in logarithmic scale.

This is a particularity of the time series analysis which is in principle one-dimensional, thus when the source is shifted from one pixel to another this is interpreted of the oscillatory power.

In Figure 2 we plot the oscillatory power present in the 1700 Å data of AR 11385. This region showed much more oscillatory power than in either of 11386 or 11387. Figure 2 was formed by integrating the oscillatory power over the whole analyzed area. All strong flares in emerging AR 11387 seem well correlated with an increase in the oscillatory signal of 90% above AR 11385 (Table 1; flares c, e, g ). However, there is no oscillatory increase from flares originating from AR 11386 (Table 1, flares b and d).

Table 1. List of Flares Used in This Study

| | Flare Coordinates | Type of Flare | Region | Time |
|---|---|---|---|---|
| a | S22 N13 | C1.6 | emerging AR 11387 | 4:47 |
| b | S18 E51 | C5.2 | AR 11386 | 8:27 |
| c | S22 N13 | C5.5 | emerging AR 11387 | 8:49 |
| d | S18 E 51 | C1.1 | AR 11386 | 11:04 |
| e | S22 N13 | C8.4 | emerging AR 11387 | 11:20 |
| f | S22 N13 | C2.2 | emerging AR 11387 | 16:55 |
| g | S22 N13 | M4.0 | emerging AR 11387 | 18:11 |

Figure 3. shows oscillatory power detected in AR 11386, the source of two of the observed flares(Table 1, flares b and d). Flares that occurred in the region itself did not cause a significant response response in oscillatory behavior of the region. A smaller response to the 3 strongest flares, C5, C8, and M4, from emerging region AR 11387 is evident(Table 1; flares c, e, g ). There is also a significant delayed response to the flare C8.4 (Table 1; flare e) occurring at 11:20 from the region AR 11387 which might indicate an additional disturbance reaching the region later in the evolution of the flare. A similar delayed response is registered from the strongest flare M4 from region AR 11387 (Table 1; flare g) , but this oscillatory power is much less than from weaker C8.5 flare (Table 1; flare e). The third prominent group of active regions combining AR 11384 and parts of AR 11383 did not produce any strong flares itself (Fig. 4). However this area did exhibit an oscillatory response to 4 flares (Table 1; flares a, c,e, g) from emerging region AR 11387. The response to these flares was much weaker than the response in region AR 11385, but stronger than the response of AR

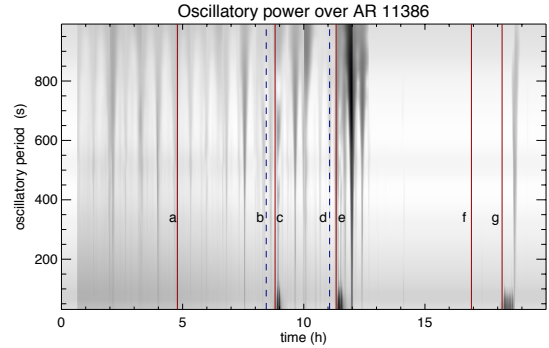

Fig. 3.— Logarithmic power-plot of detected oscillations that occurred in FOV encompassing AR 11386. Vertical lines mark times when flares occurred. Markings are same as in Fig. 2 .

11386. Here too we can see an additional delayed response to the strongest g flare class M4 from region AR 11387. This region did not responded to C2.2 flare (Table 1; flare f) from region AR 11387 nor flares from AR 11386.

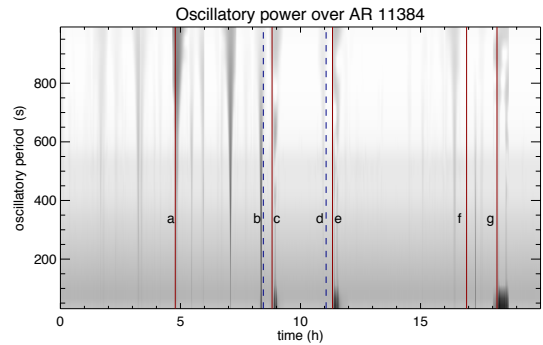

Fig. 4.— Logarithmic power plot of detected oscillations in FOV encompassing AR 11384 and AR 11383. Markings are same as for Figs. 2 and 3.

The connection between the onset of a flare and the oscillatory response has two possible origins; one is that the magnetic field carries energy between regions, the second is that a globally propagating wave (e.g., a Moreton wave) originating from the flare causes a oscillation in another region. An analysis of the quiet Sun between the regions may discriminate between these possibilities. We analyzed two different regions of quiet Sun, one in the southern hemisphere between regions AR 11386 and AR 11385 and one in the northern hemisphere between AR 11383 and west limb (Fig. 1, squares marked with QS1 and QS2). Both regions showed a similar oscillatory signal (Fig.5).

no flares and during the M4 flare from region AR 11387 (Table 1; flare g) Clearly the oscillatory behavior in the region changes when a flare occurs outside that region. We

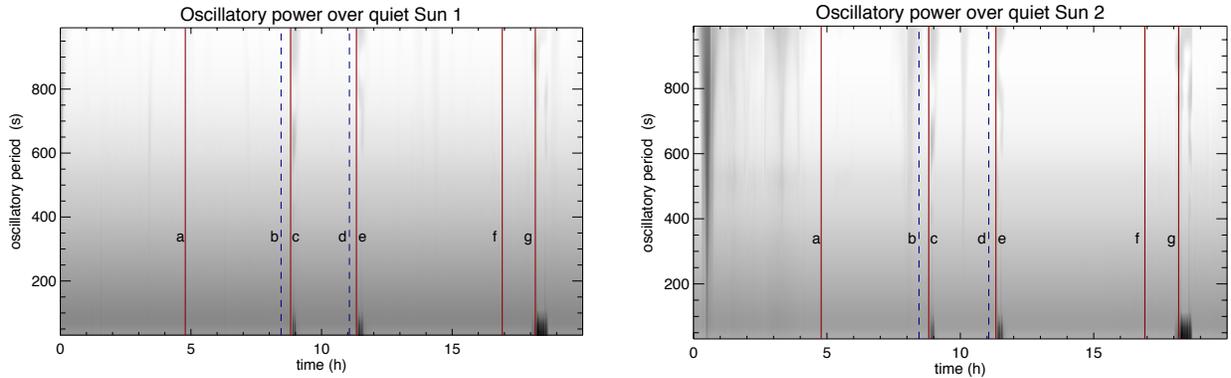

Fig. 5.— Logarithmic power-plot of detected oscillations in two different quiet sun areas. Areas are located on opposite sides of the disc equator. Region 1 is between AR 11386 and AR 11385, while region 2 is between AR 11383 and west limb. Markings on figures are same as for Figs. 2,3, and 4.

There is an oscillatory signal correlated with the flares C5, C8, and M4 from AR 11387 (Table 1; flares c, e, g), however this is significantly weaker than the oscillatory power in the active regions. There is no oscillatory power connected to any smaller flare, nor to any flare from AR 11386. The quiet Sun region located in the northern hemisphere showed a stronger response to the M4 flare (Table 1; flare g) than the region in the southern hemisphere. However, the northern quiet Sun region also had more oscillatory activity than the southern region.

summed the number of the cycles over the whole time when flares in region were not present, and then normalized to the duration of the flare. Prior to the flare, most oscillations showed a short duration, with the most numerous oscillations around 150 s period (Fig. 6. left panel). This situation changes dramatically during the flare. The duration of oscillations with the periods of 150s and below increases by several orders of magnitude, from near zero values to the order of a thousand registered wave-trains that last over 80 cycles. This was repeated in

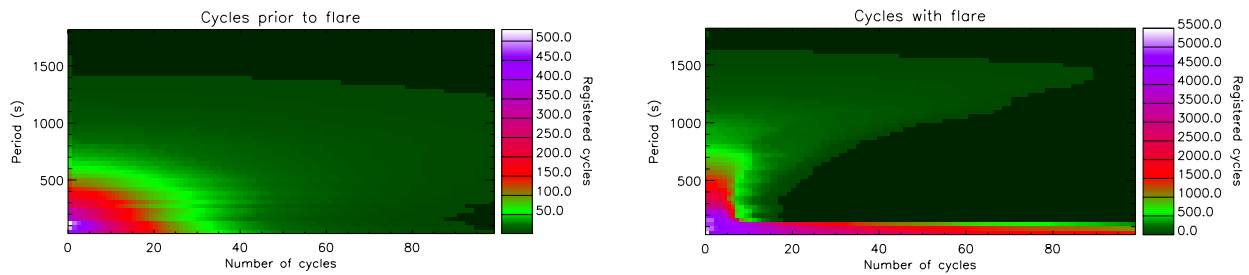

Fig. 6.— Count of cycles completed by registered oscillations. The left panel shows the number of cycles in AR 11385 prior to the flare, wheras the right panel shows the number of cycles in AR11385 during the M4 flare from AR11387.

Figure 6 shows a comparison of the duration of the oscillations in AR 11385 when there is each flare.

Figure 7 shows the time delay over different distances from the flare locations. Distances were calculated from the center of each region. The panel shows the flare location in emerging region AR 11387, while the bottom one the flare location in region AR11386. Oscillatory reaction does not peak in all frequency ranges simultaneously. In Figure 7 we present only the first peak in any of the frequencies. If the flare did not produce a reaction in some regions (as for flares b and f) the time delay was registered as 0. Although there is a variation in time response, there is no linear relationship for either case. The averaged linear Pearson correlation coefficient is only 0.0533, indicating that there is no linear relation between distance of the area from the flare location and the time delay between flare and oscillatory reaction.

magnetic activity (by removing the quiet Sun regions from the correlation calculations) allows us to approximate similar magnetic strengths in the regions and thus reduce the influence introduced by the different strengths of the magnetic field. The resulting averaged linear Pearson correlation coefficient is 0.7987, indicating that when we control for difference in magnetic field, there might be some linear dependence between distance and the power of the oscillatory reaction. The fact that the same averaged linear Pearson correlation coefficient drops to the 0.4641 when we do not control for the magnetic difference indicates that the relation between power of the reaction and the magnetic strength might be stronger than the relationship between distance between flare location and the location of the oscillatory response.

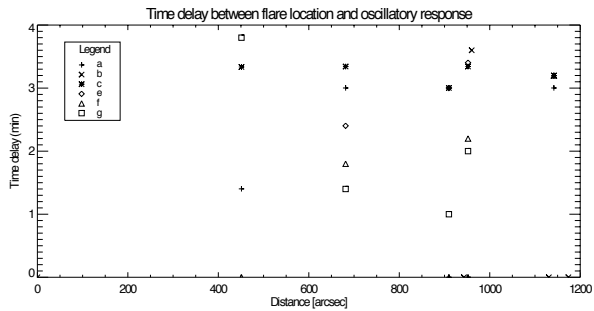

*Fig. 7.— The time delay between flare locations and the oscillatory response is shown against distance to the analyzed regions. If there was no reaction, the time delay is set to 0. Flare d is not shown because it did not cause reaction in any of the observed regions.*

The registered power of the oscillatory reaction was calculated by integrating power over the region and over the observed frequency range. If there was no reaction, the registered power is same as the background power of the region. The top panel shows the flare location in emerging region AR 11387, while bottom one shows the flare location in region AR 11386. There is no clear relationship visible from the Fig. 8, most likely due to the difference in magnetic strength of the regions. Controlling for these difference in

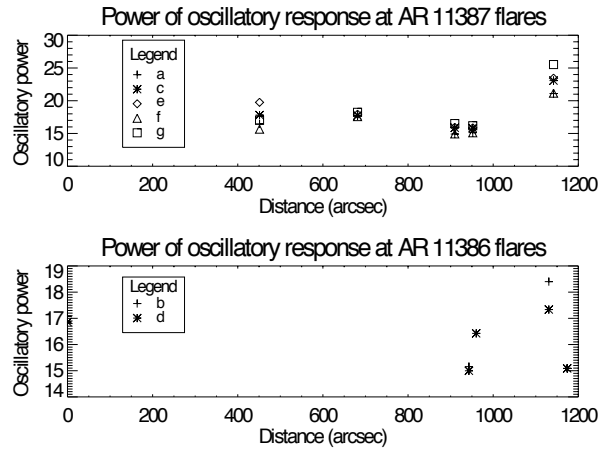

*Fig. 8.— The logarithm of integrated power of oscillatory response is presented against distance of the analyses regions. Top panel shows the emerging active region AR387 with five flares (a, c, e, f, and g). The bottom panel represents active region AR386 with two flares (b and d). If there was no reaction, the power presented reflects the oscillatory power of the omnipresent oscillatory modes.*

## 4. Discussion

We show that solar flares can result in an increase in oscillatory power in the chromosphere. An increase in oscillatory power occurs in non-flaring regions of the solar disk, but not in the flaring region itself. The distance of the oscillating region from the flare plays a significant role, with regions similar in magnetic strength show linear correlation with distance. This finding indicates that there is a possible large scale connection between different active regions. The nature of this connection may be either through the magnetic field or a globally propagating wave. Although we observed a temporal difference, this difference is not linear with the distance (Fig.7, Table 2). With our temporal resolution it appears that the oscillatory response happens in all observed regions at the similar time, irrespective of the distance of the region from the flare location. This lack of the clear, linear, temporal difference between regions in response regardless to their distance of the flare location indicates that the magnetic field connections may play a more significant role than a wave in the transfer of the disturbance.

Table 2. Linear Pearson correlation coefficients

| Flare | Time delay of reaction | Integrated oscillatory power |
|---|---|---|
|  | (all regions) | (active regions) |
| a | -0.0127 | 0.9517 |
| b | 0.1377 | 0.5695 |
| c | -0.4431 | 0.9448 |
| d | - | 0.2893 |
| e | -0.0915 | 0.9387 |
| f | 0.6778 | 0.9515 |
| g | 0.0519 | 0.9450 |

The oscillatory reactions of filaments caused by flare are thought due to Moreton waves (Balasubramaniam et al, 2007, 2010, Asai et al. 2012, Li & Zhang 2012, Jackiewicz & Balasubramamniam, 2013). If we assume that a Moreton wave propagates ≈ 1000 km/s (Moreton 1960) then such a wave caused by strongest flares would reach AR 11386 in ≈ 11 minutes, AR 11385 in ≈ 3 minutes, and QS2 in ≈ 2 minutes. While the reaction times for region AR 11385 are in agreement of the arrival of a possible Moreton wave, all other regions react at exactly same time as region AR 11385, thus either reacting earlier than they should (AR 11386, AR 11384 and QS1) or later then they should (QS2). This test seemingly eliminates a Moreton wave as the possible cause of reaction to all regions except for AR 11385 (Fig. 9). Fig. 9 shows the reactions of the 4 different regions to the C5.5 flare from emerging region AR 11387. The solid blue line is a the time of the flare, while the dashed red line is at the calculated arrival of the a possible wave to the region. All regions react at the exact same time to the flares, regardless of their distance to the flares. This simultaneous reaction indicates that a non-wave mechanism may be more likely. We speculate that the disturbance is transferred through the large scale magnetic field. However, with only 3 active regions and 7 flares analyzed, additional studies are necessary to confirm this.

The indication that magnetic field plays a significant role is first emphasized with the situation observed in AR 11386 where we have a strong delayed response from a C8.8 but a much smaller delayed response from an M4 flare. Although both flares occurred at the same distance from the region, we can speculate that during the time between the C8.8 and the M4 the magnetic field has changed to reduce the ability of the field to transfer the oscillatory power. The calculation of the linear Pearson correlation coefficient between distances and strength of oscillatory reaction indicate a linear relationship only when we perform the control for the differences in the magnetics strengths of the regions.

Quiet Sun regions also react to the strongest flares only, but with a much smaller increase of the oscillatory power. It is possible that either some kind of globally propagating wave is only accompanying the strongest flares, or only the disturbance that stronger flares cause in magnetic field is transferred through the whole global solar field. Another possibility is that the reaction is a consequence of an unobserved internal mechanism. Clearly further examples are necessary, especially because there is no apparent time delay between the response in the quiet Sun and the response in either of active regions.

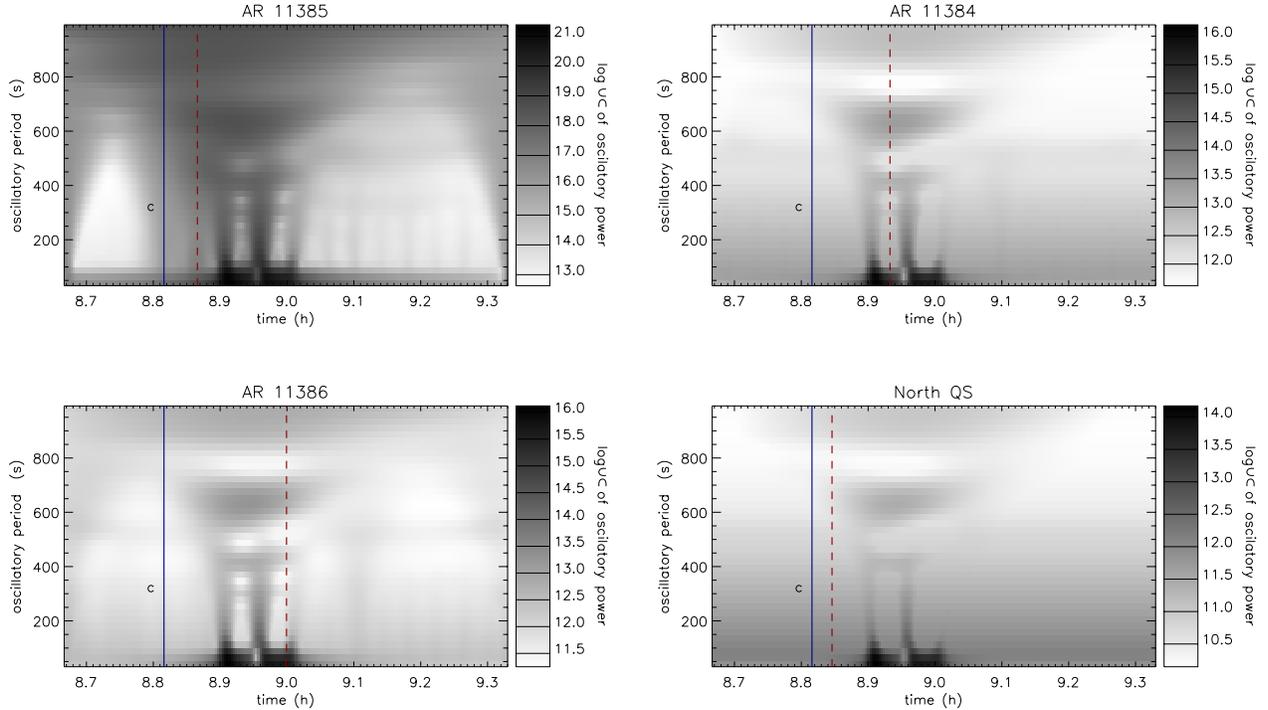

Fig. 9.— Zoom in of the time evolution of the detected oscillatory power for 4 different areas. The solid blue line is at the time of the flare, while the dashed red line is the time of a Moreton wave of velocity ≈ 1000 kms$^{-1}$.

Oscillations tend to form longer wave-trains during the time of the flare than prior to it. This prolonged duration of uninterrupted oscillatory cycles implies that more energy is transferred with longer wave-trains, especially with short-period oscillations. From the theoretical standpoint short-period oscillations should provide energy through dissipation (Vigeesh et al 2009, Shelyag et al. 2009). Thus we can speculate that there is an indication that this reaction to flares from different locations tends to redistribute part of the energy released by the flare to the other active regions.

We have shown that flares do cause an increase in oscillatory flux in locations distant from the flare. This increase probably depends on the number of magnetic flux tubes present and distance from the flare location. In future work we will quantify exactly how this disturbance is transferred to the different solar regions and how this disturbance effects the total energy balance of the solar atmosphere.

*We would like to thank anonymous referee for the helpful comments. We gratefully acknowledge support of NSF (PAARE AST-0849986), NASA (EPSCOR NNX09AP76A) and NASA (NNX13AE03G).*